\documentclass[twocolumn,showpacs,preprintnumbers,amsmath,amssymb]{revtex4}
\usepackage{graphicx}% Include figure files
\usepackage{dcolumn}% Align table columns on decimal point
\usepackage{bm}% bold math
\begin{document}
\title{Umklapp Frustration and Field-theoretic Approach to Superconductivity}
\date{\today}
\author{X. H. Zheng and D. G. Walmsley}
\affiliation{Dept.\ Phys.\ Astronomy, Queen's Univ.\ Belfast, BT7
1NN, N.\ Ireland} \email{xhz@qub.ac.uk}

\begin{abstract}
The formula of Gell-Mann and Low can be applied to both the Stark
effect and superconductivity. The standard version of the
field-theoretic approach fits the Stark effect, because in this
version electrons have identical initial and end states, so that
the energy of each and every electron orbit acquires a shift in
the same direction. A preliminary alternative version of the
field-theoretic approach is introduced, in which the electrons
have different initial and end states, to accommodate
superconductivity.  Consequently the energy of the electrons
acquires a random shift, which will cancel macroscopically, unless
the electrons are paired. It also becomes apparent that
superconductivity is frustrated when normal and umklapp scattering
coexist.
\end{abstract}

 \maketitle

\section{introduction}\label{sec:A}
\begin{figure}
\resizebox{8cm}{!}{\includegraphics{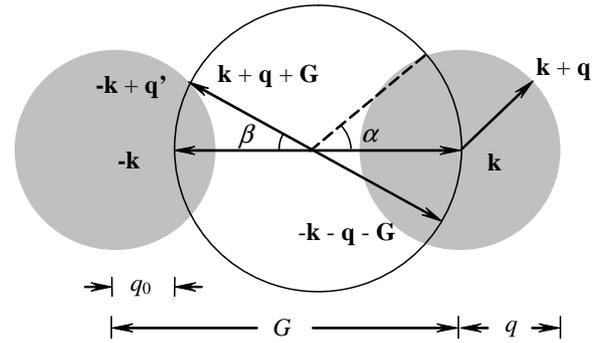}}
\caption{\label{fig:fig1} A spherical Fermi surface (open circle)
and a pair of electrons in states $\bf k$ and $\bf -k$. This pair
is scattered into state $\bf k + q + G$ and $\bf-k - q - G$ in the
umklapp process, $\bf q$ being the phonon wavevector, $\bf G$
reciprocal lattice vector. The shaded circles represent two
identical spherical phonon zones. We use angles $\alpha$ and
$\beta$ to measure the size of the sections of the Fermi surface
being intersected by these phonon zones.  We have $\alpha > \beta$
because a larger section of the Fermi surface is intersected by
the phonon sphere centered at this surface (the other sphere is
centered above the Fermi surface at height $q_0$). As a result,
the electron state $\bf-k - q - G$ must lie inside the phonon
sphere centered at $\bf k$, where all the states are involved in
normal scattering.}
\end{figure}

In umklapp scattering a Cooper pair may contest a state which is
the destination of another pair in normal scattering
(FIG.~\ref{fig:fig1}). Consequently the contribution to
superconductivity is frustrated~\cite{ZW1}. We find that on
average only 15\% of phonons are involved in pairing electrons in
12 common superconductive metals. Were the frustration effect of
umklapp scattering lifted, metals would be superconductive at
remarkably high temperatures, for example 980 K in the case of
tin~\cite{ZW2}. This result is found from the original approach of
Bardeen, Cooper and Schrieffer (BCS), based on the method of
variation~\cite{BCS}. It is natural to ask if the same result
could be found from the standard field-theoretic approach, based
on the method of the Green function~\cite{Abrikosov, Schrieffer}.

%%%%%%%%%%%%%%%%%%%%%%%%%%%%%%%%%%%%%%%%%%%%%%%%%%%%%%%%%%%%%%%%%
%%%%%%%%%%%%%%%%%%%%%  FIG.1 is here.   %%%%%%%%%%%%%%%%%%%%%%%%%

The Green function method can be highly sophisticated.  As a first
attempt, we follow a relatively simple approach which, to our
knowledge, was pioneered by Hubbard~\cite{Hubbard} to describe
collective motions in terms of many-body perturbation theory, with
many features common to the standard field-theoretic approach. For
example both approaches start with the formula of Gell-Mann and
Low~\cite{Gell-Mann}, which involves an expansion of the
scattering matrix or $s$-matrix.  In both approaches the treatment
of Wick is applied in order to reduce the operator chain into a
$c$-number~\cite{Wick}.  The diagram technique is also used in
both approaches to identify the relevant physical process and
facilitate calculation, etc.  Detailed knowledge about one
approach will certainly help us to understand the other.

In a field-theoretic approach it is customary to start with an
ensemble of free particles.  Then the interaction is switched on
adiabatically in time.  According to Gell-Mann and Low an
eigen-state of the free particles evolves adiabatically and
eventually becomes an eigen-state of the ensemble with
interaction. Apparently this approach is applicable to say the
Stark effect, which can be literally switched on adiabatically,
and is characterized by either of the following chains of events:
\begin{eqnarray}
\begin{array}{c}({\bf k}, \sigma) \rightarrow ({\bf k + q}, \sigma)
\rightarrow ({\bf k}, \sigma)\\\\
({\bf k + q}, \sigma) \rightarrow ({\bf k}, \sigma) \rightarrow
({\bf k + q}, \sigma)\end{array}\label{eq:A1}
\end{eqnarray}
$\bf k$ and $\bf q$ being the wave-vectors of the electron and
photon respectively, $\sigma = \uparrow$ or$\downarrow$ spin.  In
the first line of Eq.~(\ref{eq:A1}) the electron absorbs a photon
from the external field and then emits it.  In the second line the
electron emits a photon and then absorbs one again.  In both cases
the electron returns to its original state, reflecting that the
external field alters the electron orbit, rather than forces the
electron to jump orbit.

We find through this simple version of the field-theoretic
approach that, when we switch on the electron-phonon interaction
in a particle ensemble, the ensemble energy changes.  Apart from
an additional factor of 2, the strength of the interaction is the
same as that in the BCS theory. The electrons always return to
their original state, a situation similar to the Stark effect, but
different from the picture of the BCS theory. Indeed, we find that
the energy change of each and every electron is of the same sign
and they do not cancel. There is no need to pair the electrons.
Consequently there can be no umklapp frustration of
superconductivity.  We emphasize we are not criticizing the work
of Hubbard: he did not employ his approach to study
superconductivity~\cite{Hubbard}.

To deal with superconductivity we adapt the approach
in~\cite{Hubbard} slightly. We alter the path of scattering so
that the electrons no longer return to their original state.
Pairing becomes necessary, otherwise there will be no change of
the ensemble energy.  Now the process is characterized by the
following event:
\begin{eqnarray}
\begin{array}{c}({\bf k}, \uparrow) \rightarrow ({\bf k + q},
\uparrow)\\\\
({\bf-k}, \downarrow) \rightarrow ({\bf-k - q}, \downarrow)
\end{array}\label{eq:A2}
\end{eqnarray}
Here the two electrons jump orbits after exchanging virtual
phonons. The energy of the electron ensemble is lowered, not
because some slowly introduced interaction alters the energy of
the electron orbits. Rather, when the two electrons jump orbits in
a coherent manner, they are in a configuration of a lower energy.
We also find umklapp frustration.

This communication is arranged as follows.  In Section~\ref{sec:B}
we introduce the Hamiltonian of the electron-phonon system.  In
Sections~\ref{sec:C}~-~\ref{sec:F} we derive the energy shift
according to the scenario in Eq.~(\ref{eq:A1}). In
Section~\ref{sec:G} we derive the energy shift according to the
scenario in Eq.~(\ref{eq:A2}). In Section~\ref{sec:H} we show that
pairing in necessary for this second scenario. In
Section~\ref{sec:I} we reveal umklapp frustration to
superconductivity.  In Section~\ref{sec:II} we discuss Coulomb
repulsion.  In Section~\ref{sec:J} we outline a possible path to a
self-consistent solution of our alternative field-theoretic
approach.  In Section~\ref{sec:K} we comment on the standard
field-theoretic approach.  We give a brief conclusion in
Section~\ref{sec:L}.

\section{electron-phonon system}\label{sec:B}
Unless stated otherwise, we always use the Schr\"odinger
representation.  We separate the Hamiltonian into two parts:
\begin{eqnarray}
H = H_0 + H' \label{eq:B1}
\end{eqnarray}
where $H_0$ and $H'$ are Hamiltonians of free particles and
interaction, respectively, of which the later will be considered
as perturbation. In an electron-phonon system we have
\begin{eqnarray}
H_0 = H_e + H_p \label{eq:B2}
\end{eqnarray}
where $H_e$ and $H_p$ are electron and phonon Hamiltonian
respectively.  In second quantization we have
\begin{eqnarray}
&&H_e = \sum_{{\bf k},\sigma}\epsilon_0({\bf k})a^\dag_{{\bf
k},\sigma}a_{{\bf k},\sigma}, \label{eq:B3}\\
&&H_p = \sum_{{\bf q},l}\hbar\omega_l({\bf q})\left(c^\dag_{{\bf
q},l}c_{{\bf q},l} + 1/2\right),\label{eq:B4}
\end{eqnarray}
$a^\dag$ and $a$ being the electron generation and destruction
operators, $\epsilon_0$ and $\bf k$ electron energy and wave
vector in the absence of interaction, $c^\dag$ and $c$ phonon
generation and destruction operators, $\omega$ and $\bf q$ phonon
frequency and wave vector and $l$ identifies phonon polarization.
The interaction Hamiltonian turns out to be
\begin{eqnarray}
H'\!\!=\!\!\sum_{{\bf k}, \sigma, {\bf q}, l}\mathcal{M}_{{\bf q},
l}c_{{\bf q}, l}a^\dag_{{\bf k + q}, \sigma}a_{{\bf k}, \sigma}+
h.c.\label{eq:B5}
\end{eqnarray}
$h.c.$ stands for Hermitian conjugate.  Here we neglect the slight
dependence of the matrix element $\mathcal M$ on $\bf
k$~\cite{BCS}.  Note that in Eq.~(\ref{eq:B5}) an electron has the
same spin before and after being scattered by a phonon.

\section{Identical initial and end states}\label{sec:C}
Gell-Mann and Low proved the following important
formula~\cite{Hubbard,Gell-Mann}:
\begin{equation}
|\psi\rangle =
S(t)|\psi_0\rangle/\langle\psi_0|S(0)|\psi_0\rangle.\label{eq:C1}
\end{equation}
where $|\psi\rangle$ and $|\psi_0\rangle$ satisfy
\begin{eqnarray}
\begin{array}{lll}H|\psi\rangle = \epsilon|\psi\rangle&\mbox{and}&
H_0|\psi_0\rangle = \epsilon_0|\psi_0\rangle,\end{array}
\label{eq:C2}
\end{eqnarray}
whereas
\begin{eqnarray}
S(t) = S_0(t) + S_1(t) + S_2(t) + ...\label{eq:C3}
\end{eqnarray}
with $S_0(t) = 1$,
\begin{eqnarray}
S_n(t) &=& \frac{1}{(i\hbar)^n}
\int_{-\infty}^t\!\!\!dt_1\int_{-\infty}^{t_1}\!\!\!dt_2\nonumber\\
&...&\int_{-\infty}^{t_{n-1}}\!\!\!dt_n
H'_\alpha(t_1)H'_\alpha(t_2)...H'_\alpha(t_n)\label{eq:C4}
\end{eqnarray}
which is known as the $s$-matrix (scattering matrix),
\begin{equation}
H'_\alpha(t) = \lim_{\alpha\rightarrow+
0}e^{-H_0t/i\hbar}H'e^{H_0t/i\hbar}e^{\alpha t}.\label{eq:C5}
\end{equation}
For clarity we will integrate Eq.~(\ref{eq:C4}) directly, instead
of the usual practice in the literature of replacing $t_1$, $t_2$,
..., $t_{n-1}$ with $t$ in the multiple integration of time via
the use of the chronological operator.

It is apparent from Eq.~(\ref{eq:C1}) that, when $t = 0$, we have
$\langle\psi_0|\psi\rangle = 1$ and this leads through
Eqs.~(\ref{eq:B1}) and (\ref{eq:C2}) to
\begin{eqnarray}
\epsilon - \epsilon_0 &=& (\epsilon -
\epsilon_0)\langle\psi_0|\psi\rangle\nonumber\\
&=& \langle\psi_0|H|\psi\rangle -
\langle\psi_0|H_0|\psi\rangle\nonumber\\
&=& \langle\psi_0|H'|\psi\rangle \label{eq:C6}
\end{eqnarray}
Combining Eqs.~(\ref{eq:C1}) and (\ref{eq:C6}) we find
\begin{eqnarray}
\epsilon - \epsilon_0 = \frac{\langle\psi_0|H'S(0)|\psi_0\rangle}
{\langle\psi_0|S(0)|\psi_0\rangle} \label{eq:C7}
\end{eqnarray}
or $\epsilon - \epsilon_0 = \epsilon_1 + \epsilon_2 + ...$ with
\begin{eqnarray}
\epsilon_n =\frac{\langle\psi_0|H'S_n(0)|\psi_0\rangle}
{\langle\psi_0|S(0)|\psi_0\rangle}\label{eq:C8}
\end{eqnarray}
Note that in Eqs.~(\ref{eq:C7}) and (\ref{eq:C8}) the initial and
end states are identical.  This is why Eq.~(\ref{eq:C8}) vanishes
with $S_0(0) = 1$: the phonon field in the bra is altered by the
phonon operator in $H'$ and becomes orthogonal to the phonon field
in the ket (or vice versa). Apparently, $\epsilon_1$ is associated
with the scenarios of scattering in Eq.~(\ref{eq:A1}), where the
particle always returns to its initial state.

\section{S-matrix evaluation}\label{sec:E}
We seek the matrix element of $H'_\alpha$ in the Schr\"odinger
representation.  To this end we have to evaluate
\begin{eqnarray}
\begin{array}{c}\langle0|a_{\bf k + q, \sigma}H'_\alpha c^\dag_{{\bf q},
l}a^\dag_{\bf k, \sigma}|0\rangle\\\\
\langle0|c_{{\bf q}, l}a_{\bf k, \sigma}H'_\alpha a^\dag_{\bf k +
q, \sigma}|0\rangle\end{array}\label{eq:E1}
\end{eqnarray}
$|0\rangle$ being the electron and phonon vacuum, which tell us
the probability of any event in Eqs.~(\ref{eq:A1}) and
(\ref{eq:A2}). We have
\begin{eqnarray}
&&e^{H_0t/i\hbar}\;c^\dag_{{\bf q}, l}a^\dag_{{\bf
k},\sigma}|0\rangle\nonumber\\
&&= \sum_{n = 0}^\infty\frac{(t/i\hbar)^n}{n!}H_0^n\;c^\dag_{{\bf
q}, l}a^\dag_{{\bf k},\sigma}|0\rangle\nonumber\\
&&= \sum_{n = 0}^\infty\frac{(t/i\hbar)^n}{n!}\big[\epsilon_0({\bf
k}) + \hbar\omega_l({\bf q})\big]^nc^\dag_{{\bf q},
l}a^\dag_{{\bf k},\sigma}|0\rangle\nonumber\\
&&=\exp\Big\{\big[\epsilon_0({\bf k}) + \hbar\omega_l({\bf
q})\big]t/i\hbar\,\Big\}\,c^\dag_{{\bf q},l}a^\dag_{{\bf
k},\sigma}|0\rangle\label{eq:E2}
\end{eqnarray}
Similarly we have
\begin{eqnarray}
&&e^{H_0t/i\hbar}\;a^\dag_{{\bf k + q},\sigma}|0\rangle\nonumber\\
&&=\exp\big[\epsilon_0({\bf k + q})t/i\hbar\big]\,a^\dag_{{\bf k +
q},\sigma}|0\rangle\label{eq:E3}
\end{eqnarray}
Combining Eqs.~(\ref{eq:B5}), (\ref{eq:C5}) and
(\ref{eq:E1}-\ref{eq:E3}), we find
\begin{eqnarray}
H'_\alpha&=&\!\sum_{{\bf k},\sigma,{\bf q},l}\mathcal{M}_{{\bf
q},l}\,c_{{\bf q},l}\,a^\dag_{{\bf k+q},\sigma}a_{{\bf
k},\sigma}\nonumber\\
&\times&\exp\Big\{\big[\epsilon_0({\bf k}) + \hbar\omega_l({\bf
q}) - \epsilon_0({\bf k+q})\big]t/i\hbar\Big\}\nonumber\\
&+&\!\sum_{{\bf k},\sigma,{\bf q},l}\mathcal{M}_{{\bf
q},l}\,c^\dag_{{\bf q},l}a^\dag_{{\bf k},\sigma}a_{{\bf
k + q},\sigma}\nonumber\\
&\times&\exp\Big\{\big[\epsilon_0({\bf k + q}) - \epsilon_0({\bf
k}) - \hbar\omega_l({\bf
q})\big]t/i\hbar\Big\}\;\;\;\;\;\;\;\label{eq:E5}
\end{eqnarray}
where for simplicity we have dropped the factor $\exp(\alpha t)$
in Eq.~(\ref{eq:C5}).

In order to evaluate the integrations in Eq.~(\ref{eq:C4}) at $t =
-\infty$, we add an infinitesimal positive number $\delta$ to the
argument of the exponential functions in Eq.~(\ref{eq:E5}). This
leads to
\begin{eqnarray}
&&S_1(0) =\!\!\!\sum_{{\bf k},\sigma,{\bf q},
l}\!\!\!\mathcal{M}_{{\bf q}, l}\left[\frac{c_{{\bf q},
l}a^\dag_{{\bf k + q},\sigma}a_{{\bf k},\sigma}}{\epsilon_0({\bf
k}) + \hbar\omega_l({\bf q}) -
\epsilon_0({\bf k + q}) + i\delta}\;\;\right.\nonumber\\
&&\;\;\;\;\;\;\;\;\;\;\;\;\;\;\;\;\;\;\;\;+\left.\frac{c^\dag_{{\bf
q}, l}a^\dag_{{\bf k},\sigma}a_{{\bf k +
q},\sigma}}{\epsilon_0({\bf k + q}) - \epsilon_0({\bf k}) -
\hbar\omega_l({\bf q}) + i\delta}\right]\label{eq:E6}
\end{eqnarray}
as the $n = 1$ term of $S(0)$.

\section{Stark effect}\label{sec:F}
We substitute Eq.~(\ref{eq:E6}) into Eq.~(\ref{eq:C8}) to evaluate
$\epsilon_1$.  We find we have to evaluate the following:
\begin{eqnarray}
\langle\psi_0|c^\dag_{{\bf q'\!}, l'\!}a^\dag_{{\bf k'\!},
\sigma'}a_{{\bf k'\! + q'\!},\sigma'}c_{{\bf q}, l}a^\dag_{{\bf k
+ q} ,\sigma}a_{{\bf k}, \sigma}|\psi_0\rangle\label{eq:F2}\\
\langle\psi_0|c^\dag_{{\bf q'\!}, l'}a^\dag_{{\bf k'\!},
\sigma'}a_{{\bf k'\! + q'\!},\sigma'}c^\dag_{{\bf q},
l}a^\dag_{{\bf k}\sigma}a_{{\bf k + q},
\sigma}|\psi_0\rangle\label{eq:F3}\\
\langle\psi_0|c_{{\bf q'\!}, l'}a^\dag_{{\bf k'\! +
q'\!},\sigma'}a_{{\bf k'\!},\sigma'}c_{{\bf q}, l}a^\dag_{{\bf k +
q},\sigma}a_{{\bf k},\sigma}|\psi_0\rangle\label{eq:F4}\\
\langle\psi_0|c_{{\bf q'\!}, l'\!}a^\dag_{{\bf k'\! +
q'\!},\sigma'}a_{{\bf k'\!},\sigma'}c^\dag_{{\bf q},
l}a^\dag_{{\bf k}, \sigma}a_{{\bf k + q},
\sigma}|\psi_0\rangle\label{eq:F5}
\end{eqnarray}
In Wick's treatment Expressions~(\ref{eq:F2} - \ref{eq:F5}) are
evaluated via permutation of the operators.  Description of this
process in the literature is often in some special terminology,
such as `$N$-product', `$T$-product', `pairing', etc.  Essentially
we have to move some of the operators to the left to apply on the
bra, move others to the right to apply on the ket. The transformed
bra and ket must be symmetrical (that is the bra is the Hermitian
conjugate of the ket and vice versa), otherwise their product will
vanish. The sign of this product depends on the nature of the
operators and the number of permutations we have made. Since we
have two phonon generation operators in Expression~(\ref{eq:F3}),
the bra and ket can never be transformed symmetrically, so that
this expression has to vanish. For the same reason
Expression~(\ref{eq:F4}) also has to vanish.

In Expressions~(\ref{eq:F2}) and (\ref{eq:F5}) the bra and ket may
be transformed symmetrically when the phonon operators have the
same momentum and polarization.  This leads to a $c$-number
$\delta_{\bf q, q'}\delta_{l, l'}$, which in turn leads to
\begin{eqnarray}
\epsilon_1 &=&\!\!\!\!\sum_{{\bf k, k'\!}, \sigma,
\sigma'}\sum_{{\bf q}, l}\frac{\mathcal{M}_{{\bf q},
l}^2}{\langle\psi_0|S(0)|\psi_0\rangle}\nonumber\\
&\times&\left[\frac{\langle\psi_0|a^\dag_{{\bf k'\!},
\sigma'}a_{{\bf k'\! + q}, \sigma'}a^\dag_{{\bf k + q},
\sigma}a_{{\bf k}, \sigma}|\psi_0\rangle}{\epsilon_0({\bf k}) +
\hbar\omega_l({\bf q})
- \epsilon_0({\bf k + q}) + i\delta}\right.\nonumber\\
&+&\left.\frac{\langle\psi_0|a^\dag_{{\bf k'\! + q},
\sigma'}a_{{\bf k'}, \sigma'}a^\dag_{{\bf k}, \sigma}a_{{\bf k +
q}, \sigma}|\psi_0\rangle}{\epsilon_0({\bf k + q}) -
\epsilon_0({\bf k}) - \hbar\omega_l({\bf q}) + i\delta}\right]
\label{eq:F6}
\end{eqnarray}
Furthermore, in order to transform the bras and kets in
Eq.~(\ref{eq:F6}) symmetrically, we have to let $\bf k' = k$,
$\sigma' = \sigma$, in accordance with the scenario in
Eq.~(\ref{eq:A1}). We find
\begin{eqnarray}
\epsilon_1 = -2\sum_{\bf k}\sum_{\bf q}\frac{V_{\bf k,
q}}{\langle\psi_0|S(0)|\psi_0\rangle} \label{eq:F7}
\end{eqnarray}
where the factor 2 arises when we sum spin,
\begin{eqnarray}
V_{\bf k, q} = \sum_{l}\frac{2\hbar\omega_l({\bf
q})\mathcal{M}_{{\bf q}, l}^2}{\big[\hbar\omega_l({\bf
q})\big]^2\!\!- \big[\epsilon_0({\bf k + q}) - \epsilon_0({\bf
k})\big]^2} \label{eq:F8}
\end{eqnarray}
which is the same as the matrix element of the BCS reduced
Hamiltonian~\cite{ZW2}.  Note that in the above equation we have
dropped the infinitesimal number $\delta$ in Eq.~(\ref{eq:E6})
since it has no numerical effect on $V$.  In Eq.~(\ref{eq:F7}) all
the terms in $\bf k$ have the same sign, reflecting the nature of
say the Stark effect, rather than superconductivity.

\section{Different initial and end states}\label{sec:G}
According to Eq.~(\ref{eq:B1}) we have
\begin{eqnarray}
\epsilon\langle\psi|\psi\rangle = \langle\psi|H|\psi\rangle=
\langle\psi|H_0|\psi\rangle +
\langle\psi|H'|\psi\rangle\label{eq:G1}
\end{eqnarray}
By using Eq.~(\ref{eq:C1}) we find
\begin{eqnarray}
\langle\psi|H'|\psi\rangle =
\frac{\langle\psi|H'S(0)|\psi_0\rangle}
{\langle\psi_0|S(0)|\psi_0\rangle}\label{eq:G2}
\end{eqnarray}
Similarly we find
\begin{eqnarray}
\langle\psi|H_0|\psi\rangle &=&
\frac{\langle\psi|S(0)H_0|\psi_0\rangle}
{\langle\psi_0|S(0)|\psi_0\rangle}\nonumber\\
&+&\frac{\langle\psi|H_0S(0) - S(0)H_0|\psi_0\rangle}
{\langle\psi_0|S(0)|\psi_0\rangle}\label{eq:G3}
\end{eqnarray}
where
\begin{eqnarray}
\frac{\langle\psi|S(0)H_0|\psi_0\rangle}
{\langle\psi_0|S(0)|\psi_0\rangle} = \epsilon_0
\frac{\langle\psi|S(0)|\psi_0\rangle}
{\langle\psi_0|S(0)|\psi_0\rangle} = \epsilon_0
\langle\psi|\psi\rangle\label{eq:G4}
\end{eqnarray}
Combining Eqs.~(\ref{eq:G1}-\ref{eq:G4}), we find
\begin{eqnarray}
(\epsilon - \epsilon_0)\langle\psi|\psi\rangle &=&
\frac{\langle\psi|H_0S(0) - S(0)H_0|\psi_0\rangle}
{\langle\psi_0|S(0)|\psi_0\rangle}\nonumber\\
&+& \epsilon_1 + \epsilon_2 + ...\label{eq:G5}
\end{eqnarray}
with
\begin{eqnarray}
\epsilon_n = \frac{\langle\psi|H'S_n(0)|\psi_0\rangle}
{\langle\psi_0|S(0)|\psi_0\rangle}\label{eq:G6}
\end{eqnarray}
which resembles Eq.~(\ref{eq:C8}) closely, except that in
Eq.~(\ref{eq:G6}) the end state, $|\psi\rangle$, is different from
the initial state, $|\psi_0\rangle$. However we assume virtual
phonons, that is the phonon states are still identical in
$|\psi_0\rangle$ and $|\psi\rangle$, so that Eq.~(\ref{eq:G6})
vanishes with $S_0(0) = 1$, see the text below Eq.~(\ref{eq:C8}).
When $n = 1$ we have
\begin{eqnarray}
\epsilon_1&=&\!\!\!\!\sum_{{\bf k, k'\!}, \sigma,
\sigma'}\sum_{{\bf q},
l}\frac{\mathcal{M}_{{\bf q}, l}^2}
{\langle\psi_0|S(0)|\psi_0\rangle}\nonumber\\
&\times&\left[\frac{\langle\psi|a^\dag_{{\bf k'\!},
\sigma'}a_{{\bf k'\! + q}, \sigma'}a^\dag_{{\bf k + q},
\sigma}a_{{\bf k}, \sigma}|\psi_0\rangle}{\epsilon_0({\bf k}) +
\hbar\omega_l({\bf q})
- \epsilon_0({\bf k + q}) + i\delta}\right.\nonumber\\
&+&\left.\frac{\langle\psi|a^\dag_{{\bf k'\! + q}, \sigma'}a_{{\bf
k'}, \sigma'}a^\dag_{{\bf k}, \sigma}a_{{\bf k + q},
\sigma}|\psi_0\rangle}{\epsilon_0({\bf k + q}) - \epsilon_0({\bf
k}) - \hbar\omega_l({\bf q}) + i\delta}\right] \label{eq:G7}
\end{eqnarray}
which apparently is associated with the scenario in
Eq.~(\ref{eq:A2}).

Without special measures, we have $\epsilon_1 = 0$ in
Eq.~(\ref{eq:G7}). To see this, let
\begin{eqnarray}
|\psi_0\rangle = \prod_{\bf k}a^\dag_{\bf
k}|0\rangle,\;\;\;\;\;\;|\psi\rangle = \prod_{\bf k'}a^\dag_{\bf
k'}|0\rangle\label{eq:G8}
\end{eqnarray}
where for simplicity we do not generate phonons explicitly.  In
Eq.~(\ref{eq:G8}) $|\psi_0\rangle$ and $|\psi\rangle$ are
identical, except that the sequence of $a^\dag$ in $\bf k'$ is a
random reshuffle of the sequence in $\bf k$, in order to model the
chaotic nature of the electron-phonon interaction.
Eq.~(\ref{eq:G8}) allows us to transform the bras and kets in
Eq.~(\ref{eq:G7}) symmetrically after a number of permutations.
However, Eq.~(\ref{eq:G8}) also means that terms in
Eq.~(\ref{eq:G7}) will alter their signs randomly, due to the
Fermi-Dirac statistics, and cancel~\cite{BCS}.

The first term on the right hand side of Eq.~(\ref{eq:G5}) tells
us that the physics associated with $\epsilon_1, \epsilon_2, ...$
is not the sole effect of scattering.  Indeed, in order to reveal
superconductivity, Fr\"ohlich~\cite{Frohlich} introduced the
canonical transform, which cancels much of the electron-phonon
interaction.  However, we know from Eqs.~(\ref{eq:B3}),
(\ref{eq:B4}) and (\ref{eq:B5}) that terms in
\begin{eqnarray}
H_0S_1(0) - S_1(0)H_0 \label{eq:G9}
\end{eqnarray}
always have an odd number of phonon operators, which cannot
transform $|\psi\rangle$ and $|\psi_0\rangle$ in Eq.~(\ref{eq:G5})
symmetrically with respect to the phonon fields, that is we can
ignore the contribution by Expression~(\ref{eq:G9}) to $\epsilon -
\epsilon_0$, as long as we have virtual phonons, see explanation
below Eq.~(\ref{eq:G6}).

\section{Cooper pairs}\label{sec:H}
We follow BCS~\cite{BCS} to let
\begin{eqnarray}
\begin{array}{lll}{\bf k' = -k - q},&\sigma =\,\uparrow,&\sigma' =
\,\downarrow\end{array} \label{eq:H1}
\end{eqnarray}
so that Eq.~(\ref{eq:G7}) becomes
\begin{eqnarray}
\epsilon_1 = -\sum_{\bf k}\sum_{\bf q}\frac{V_{\bf k,
q}}{\langle\psi_0|S(0)|\psi_0\rangle}\,\langle\psi|b^\dag_{\bf k +
q}b_{\bf k}|\psi_0\rangle\label{eq:H2}
\end{eqnarray}
where
\begin{eqnarray}
\begin{array}{c}b_{\bf k} = a_{\bf\!-k\downarrow}
a_{\bf k\uparrow}\\\\
b^\dag_{\bf k} = a^\dag_{\bf
k\uparrow}a^\dag_{\bf\!-k\downarrow}\end{array} \label{eq:H3}
\end{eqnarray}
are the destruction and generation operators of the Cooper pairs,
which permute like Bosons~\cite{BCS}.  Clearly here we have
adopted the scenario in Eq.~(\ref{eq:A2}).  We also replace
Eq.~(\ref{eq:G8}) with
\begin{eqnarray}
|\psi_0\rangle = \prod_{\bf k}b^\dag_{\bf
k}|0\rangle,\;\;\;\;\;\;|\psi\rangle = \prod_{\bf k, q}b^\dag_{\bf
k + q}|0\rangle\label{eq:H4}
\end{eqnarray}
which leads through Eq.~(\ref{eq:H2}) to
\begin{eqnarray}
\epsilon_1 = -\sum_{\bf k}\sum_{\bf q}\frac{V_{\bf k,
q}}{\langle\psi_0|S(0)|\psi_0\rangle}\label{eq:H5}
\end{eqnarray}
$V_{\bf k, q}$ is also defined by Eq.~(\ref{eq:F8}). Note that in
Eq.~(\ref{eq:H5}) we do not sum spin, so that we do not have the
factor 2, in contrast to Eq.~(\ref{eq:F7}).

In Eq.~(\ref{eq:H4}) $\bf k + q$ in $|\psi\rangle$ runs over more
states compared with $\bf k$ in $|\psi_0\rangle$.  To understand
this we recall that we have
\begin{eqnarray}
\sum_{\bf q}V_{\bf k, q} =
\frac{1}{\Omega_D}\int_{\Omega_D}\!\!V_{\bf k, q}\,d{\bf q} =
\langle V_{\bf k, q}\rangle\label{eq:H9}
\end{eqnarray}
$\Omega_D$ being the volume of the Debye phonon sphere.
Consequently Eq.~(\ref{eq:H5}) can be written as
\begin{eqnarray}
\epsilon_1 = -\sum_{\bf k}\frac{\langle V_{\bf k,
q}\rangle}{\langle\psi_0|S(0)|\psi_0\rangle}\label{eq:H10}
\end{eqnarray}
that is $\epsilon_1$ is averaged over many end states, with the
implicit assumption that these end states can all be traced back
to the same initial state, $|\psi_0\rangle$. This situation is
opposite to that in Eq.~(\ref{eq:F7}), where the initial and end
states are the same but $\epsilon_1$ is averaged over many
intermediate states.

\section{Umklapp frustration}\label{sec:I}
We are reminded that in Eq.~(\ref{eq:G6}) we let both $\bf k$ and
$\bf k'$ run over the first electron Brillouin zone. We also let
both $\sigma$ and $\sigma'$ run over the two spins. When we pair
the electrons, we must be careful that we do not neglect or double
count any of the electron states. We may let the pairs be in
states $({\bf k}\uparrow, {\bf -k}\downarrow)$ and $({\bf
k}\downarrow, {\bf -k}\uparrow)$ and let $\bf k$ run over half the
Brillouin zone. Here we let the pairs be in the state $({\bf
k}\uparrow, {\bf -k}\downarrow)$ and let $\bf k$ run over the
whole Brillouin zone~\cite{ZW1}.  Either way we acknowledge that,
with the same wavevector $\bf k$, electrons of both spins can find
a partner with $\bf -k$ and opposite spin to form pairs. Consider
two such pairs generated by the following operators:
\begin{eqnarray}
\begin{array}{c}b^\dag_{\bf k} = a^\dag_{\bf k\uparrow}
a^\dag_{\bf\!-k\downarrow}\\\\
b^\dag_{\bf\!-k} = a^\dag_{\bf\!
-k\uparrow}a^\dag_{\bf k\downarrow}\end{array} \label{eq:I1}
\end{eqnarray}
which are in the initial state $|\psi_0\rangle$ in
Eq.~(\ref{eq:H4}). Since
\begin{eqnarray}
\langle0|b_{\bf\!-k}\,b^\dag_{\bf k}|0\rangle = \langle0|a_{\bf
k\downarrow}a_{\bf\!-k\uparrow}a^\dag_{\bf k\uparrow}a^\dag_{\bf\!
-k\downarrow}|0\rangle = 0\label{eq:I2}
\end{eqnarray}
we have two genuinely different pairs orthogonal to each other.
They are not the same pair with different names.

Now consider the end states initiated by the operators in
Eq.~(\ref{eq:I1}). In normal scattering the end states are
generated by
\begin{eqnarray}
\begin{array}{c}b^\dag_{\bf k + q} = a^\dag_{\bf k + q\uparrow}
a^\dag_{\bf\!-k - q\downarrow}\\\\
b^\dag_{\bf\!-k + q'} = a^\dag_{\bf\! -k + q'\uparrow}a^\dag_{\bf
k - q'\downarrow}\end{array}\label{eq:I3}
\end{eqnarray}
which are in $|\psi\rangle$ in Eq.~(\ref{eq:H4}). In umklapp
scattering the end states are generated by
\begin{eqnarray}
\begin{array}{c}b^\dag_{\bf k + q + G} = a^\dag_{\bf k + q + G\uparrow}
a^\dag_{\bf\!-k - q - G\downarrow}\\\\
b^\dag_{\bf\!-k + q' - G} = a^\dag_{\bf\! -k + q' -
G\uparrow}a^\dag_{\bf k - q' + G\downarrow}\end{array}
\label{eq:I4}
\end{eqnarray}
which too are in $|\psi\rangle$ in Eq.~(\ref{eq:H4}). In
FIG.~\ref{fig:fig1} we have
\begin{eqnarray}
{\bf -k + q' = k + q + G} \label{eq:I5}
\end{eqnarray}
Eqs.~(\ref{eq:I3}) and (\ref{eq:I4}) lead through (\ref{eq:I5}) to
\begin{eqnarray}
\begin{array}{c}b^\dag_{\bf k + q}b^\dag_{\bf -k + q' - G} =
b^\dag_{\bf k + q}b^\dag_{\bf k + q} = 0\\\\
b^\dag_{\bf\!-k + q'}b^\dag_{\bf k + q + G} = b^\dag_{\bf-k +
q'}b^\dag_{\bf-k + q'} = 0\end{array} \label{eq:I6}
\end{eqnarray}
which means that superconductivity is cancelled (or frustrated)
unless $\bf q$ in Eq.~(\ref{eq:H5}) is such that umklapp
scattering is not invoked.

One may ask why we do not pay attention to some other conceivable
cancellation.  For example in FIG.~\ref{fig:fig1} we may place a
second electron beside state $\bf k$ to let it compete with the
first electron for end states.  The answer lies in our initial
state, $|\psi_0\rangle$, in Eq.~(\ref{eq:H4}), where we are only
allowed to have two pairs generated by Eq.~(\ref{eq:I1}), in order
to serve the purpose of finding the energy shift for a given $\bf
k$. Indeed both $|\psi_0\rangle$ and $|\psi\rangle$ are
approximate. They are requested to serve their purpose without
self-conflict.  It is not necessary to subject them to any further
scrutiny, such as what would happen when we have the second
electron beside $\bf k$ in FIG.~\ref{fig:fig1}.

Could we go a step further to have only one pair in the initial
state? Then umklapp frustration will vanish.  Since this question
carries physical consequences, it concerns situations in physics
rather than formalisms in mathematics.  One such situation is that
either $b^\dag_{\bf k}$ or $b^\dag_{\bf\!-k}$ in Eq.~(\ref{eq:I1})
is always forbidden to generate a pair, which apparently is
unreasonable.  Another situation is that, when one pair is
scattered in accordance with the scenario in Eq.~(\ref{eq:A2}),
the other is left unchanged.  This too is unreasonable because,
apart from their spins, the two pairs generated by
Eq.~(\ref{eq:I1}) are identical.  We have no reason to treat them
differently in electron-phonon scattering, not to mention that
numerical test has already confirmed the existence of umklapp
scattering~\cite{ZW2}.

\section{Coulomb repulsion}\label{sec:II}
So far we have not discussed Coulomb repulsion, which has been
proved to have negligible effect on superconductivity, apparently
due to the situation that in a metal the Coulomb force between
electrons is in a balance, which can be toppled by even weak
attraction due to the electron-phonon interaction~\cite{ZW2}.  To
see this, we replace $H_e$ in Eqs.~(\ref{eq:B2}) and (\ref{eq:B3})
with $H_e + H_{\mbox{\scriptsize{Col}}}$, where
$H_{\mbox{\scriptsize{Col}}}$ is the Coulomb Hamiltonian, and
alter the definition of $a^\dag_{\bf k, \sigma}$ and $a_{\bf k,
\sigma}$ accordingly. Consequently we redefine $\epsilon_0$ as the
energy of the particle ensemble {\it with} Coulomb repulsion but
without the electron-phonon interaction.  We also redefine
$\epsilon$ as the ensemble energy with both the Coulomb and
electron-phonon interaction.  This has no effect on subsequent
derivations, based on the very general assumption that we have
Bloch particles, except that the numerical values of $V_{\bf k,
q}$ in Eqs.~(\ref{eq:F8}) may change slightly.

\section{Self-consistent solution}\label{sec:J}
We may follow the example of the standard field-theoretic approach
to employ the Dyson equation to find a self-consistent solution
for the superconductive energy gap~\cite{Abrikosov}.  We may
acquire a rough idea about the use of the Dyson equation from the
familiar self-consistent equation of BCS~\cite{BCS}:
\begin{eqnarray}
\Delta({\bf k}) = \sum_{\bf q}V_{\bf k, q\;}\frac{\Delta({\bf k +
q})}{2\left[\Delta^2({\bf k + q}) + \epsilon^2({\bf k +
q}))\right]^{1/2}} \label{eq:J1}
\end{eqnarray}
where $\Delta$ is the energy gap function, which replaces
$-\epsilon_1$ in Eq.~(\ref{eq:H5}).  In order to find something
like Eq.~(\ref{eq:J1}), we should be able to separate $\epsilon_1$
as a factor from $\epsilon_2, \epsilon_3, ...$ in
Eq.~(\ref{eq:G5}). We should also argue credibly that, when we add
the remainder of $\epsilon_2, \epsilon_3, ...$ together, we will
recover within a constant the original series $\epsilon_1 +
\epsilon_2 + ...$ in Eq.~(\ref{eq:G5}).  Further discussion,
however, is beyond the scope of this communication.

\section{Green function}\label{sec:K}
In the standard field-theoretic approach the superconductive
energy gap is identified from the high order terms of the Green
function, which is of the following form:
\begin{eqnarray}
G(x, x') = -i
\frac{\langle\psi_0|T\hat{\psi}(x)\hat{\psi}^\dag(x')S(0)|\psi_0\rangle}
{\langle\psi_0|S(0)|\psi_0\rangle}\label{eq:K1}
\end{eqnarray}
where $x = ({\bf r}, t)$ denotes a set of four variables,
\begin{eqnarray}
\hat\psi(x) = \sum_{\bf k, \sigma}a_{\bf k, \sigma}\psi_{\bf k,
\sigma}({\bf r})\exp[\epsilon_0({\bf k})t/i\hbar]\label{eq:K2}
\end{eqnarray}
$\psi_{\bf k, \sigma}$ being the Bloch function, and $T$ is the
chronological operator. When $t
> t'$ we have
\begin{eqnarray}
T\hat{\psi}(x)\hat{\psi}^\dag(x')S(0) =
\hat{\psi}(x)S(0)\hat{\psi}^\dag(x')\label{eq:K3}
\end{eqnarray}
otherwise, when $t\leq t'$, we have
\begin{eqnarray}
T\hat{\psi}(x)\hat{\psi}^\dag(x')S(0) =
-\hat{\psi}^\dag(x')S(0)\hat{\psi}(x)\label{eq:K4}
\end{eqnarray}
On substituting Eq.~(\ref{eq:C3}) into Eq.~(\ref{eq:K1}) we find
\begin{eqnarray}
G(x, x') = G_0(x, x') + G_1(x, x') + ...\label{eq:K5}
\end{eqnarray}
with
\begin{eqnarray}
G_n(x, x') =  -i
\frac{\langle\psi_0|T\hat{\psi}(x)\hat{\psi}^\dag(x')S_n(0)|\psi_0\rangle}
{\langle\psi_0|S(0)|\psi_0\rangle}\label{eq:K6}
\end{eqnarray}
where $S_n(0)$ is given in Eq.~(\ref{eq:C4}), $n = 0, 1, 2,
...$~\cite{Abrikosov}.

In the literature $G_0$ is known as the Green function of free
particles, of which a brief derivation is given in the Appendix,
in order to demonstrate clearly that the Green function is
evaluated with UNPAIRED particles. We know from discussion in this
communication that in $G_1$ the events are described  by the
scenario in Eq.~(\ref{eq:A1}). In the standard field-theoretical
approach the effect of superconductivity starts to become apparent
in $G_2$, which is proportional to the square of $H_\alpha'$,
similar to $\epsilon_1$ in Eq.~(\ref{eq:C8}). In $G_2$ there is
also a factor 1/2, arising from the use of $S_2(0)$ rather than
$S_1(0)$, which cancels the factor 2 in Eq.~(\ref{eq:F7}).
However, in $G_2$ the nature of events is still similar to that
described by Eq.~(\ref{eq:A1}), although the particles go through
more intermediate states before returning to their initial state,
so that the diagram technique becomes indispensable in order to
facilitate calculation.  Indeed there is no need to pair electrons
in the standard field-theoretic approach, which indicates that we
are dealing with something similar to the Stark effect rather than
superconductivity. It would be difficult to identify umklapp
frustration from the standard field-theoretic approach.

\section{conclusion}\label{sec:L}
Gell-Mann and Low devised a general formula, which is applicable
to both the Stark effect (or other similar effects) and
superconductivity. Therefore, when applying this formula, we must
know to which process we are applying it.  In the standard
field-theoretic approach the electrons always return to their
initial orbits.  The energy of each and every orbit is perturbed
in the same direction.  There is no need to pair the electrons, in
order for the process to manifest itself macroscopically.  This
communication introduces a preliminary new version of the
field-theoretic approach, where the electrons do not return to
their initial orbits. Then it becomes necessary to pair the
electrons in order to observe an energy shift of the ensemble.
Umklapp frustration emerges, which is closely associated with
pairing.  Apparently the standard field-theoretic approach fits
the nature of the Stark effect, whereas the alternative version is
more appropriate to describe superconductivity. Further work
appears to be worthwhile in order to incorporate the
sophistication of the standard field-theoretic
approach~\cite{Abrikosov} into the new version.

\section{Appendix}\label{sec:M}
When $t > t'$ we have
\begin{eqnarray}
G_0(x, x') =
-i\langle\psi_0|\hat\psi(x)\hat\psi^\dag(x')|\psi_0\rangle
\label{eq:M1}
\end{eqnarray}
where $\hat\psi$ is a weighted series of particle destruction
operators given by Eq.~(\ref{eq:K2}), in which the Bloch wave
function is of the following specific form
\begin{eqnarray}
\psi_{\bf k,\sigma}({\bf r}) = \frac{1}{\sqrt{\Omega}}\;u_{\bf k,
\sigma} \exp(i{\bf k\cdot r})\label{eq:M2}
\end{eqnarray}
$\Omega$ being the volume of the metal, $|u_{\bf k,\sigma}|^2=1$
for free particles. Combining Eqs.~(\ref{eq:K2}), (\ref{eq:M1})
and (\ref{eq:M1}), we find
\begin{eqnarray}
&&G_0(x, x') = -\frac{i}{\Omega}\sum_{\bf k,
\sigma}\langle\psi_0|a_{\bf k, \sigma}a^\dag_{\bf k,
\sigma}|\psi_0\rangle\nonumber\\
&&\times\exp\big[i{\bf k\cdot(r -
r')}\big]\exp\big[\epsilon_0({\bf k})(t - t')/i\hbar\big]
\label{eq:M3}
\end{eqnarray}
Letting $|\psi_0\rangle$ be the state of all the electrons in the
Fermi sea, which are UNPAIRED, we have
\begin{eqnarray}
\langle\psi_0|a_{\bf k, \sigma}a^\dag_{\bf k,
\sigma}|\psi_0\rangle = \left\{\begin{array}{lll}1,&&|{\bf k}|
> k_F\\\\0,&&\mbox{otherwise}\end{array}\right. \label{eq:M4}
\end{eqnarray}
$k_F$ being the Fermi wavenumber, so that
\begin{eqnarray}
&&G_0(x, x')=-\frac{i}{\Omega}\sum_{|{\bf k}|
> k_F}\exp\big[i{\bf k\cdot(r - r')}\big]\nonumber\\
&&\times\exp[\epsilon_0({\bf k})(t - t')/i\hbar] \label{eq:M5}
\end{eqnarray}
which can be written as
\begin{eqnarray}
&&G_0(x, x') = -\frac{i}{(2\pi)^3}\int d{\bf k}\,\theta\big(|{\bf
k}| - k_F\big)\nonumber\\
&&\times\exp\big[i{\bf k\cdot(r -
r')}\big]\exp\big[\epsilon_0({\bf k})(t - t')/i\hbar\big]
\label{eq:M6}
\end{eqnarray}
where
\begin{eqnarray}
\theta\big(\xi\big)=\left\{\begin{array}{lll}1,&&\xi
> 0\\\\0,&&\mbox{otherwise}\end{array}\right. \label{eq:M7}
\end{eqnarray}
When $t < t'$ we find through a similar procedure
\begin{eqnarray}
&&G_0(x, x') = -\frac{i}{(2\pi)^3}\int d{\bf k}\,\theta\big(k_F -
|{\bf k}|\big)\nonumber\\
&&\times\exp\big[i{\bf k\cdot(r -
r')}\big]\exp\big[\epsilon_0({\bf k})(t - t')/i\hbar\big]
\label{eq:M8}
\end{eqnarray}
According to Eq.~(\ref{eq:M6}), we have
\begin{eqnarray}
&&\int^{\infty}_{t'}G_0(x, x')\exp\big[i\omega(t -
t')\big]\,dt\nonumber\\
&&= \frac{\theta\big(|{\bf k}| - k_F\big)}{\omega -
\epsilon_0({\bf k})/\hbar + i\delta}\exp\big[i{\bf k\cdot(r -
r')}\big] \label{eq:M9}
\end{eqnarray}
\vspace{3mm}where we have introduced the infinitesimal number
$\delta$, in order to evaluate the integration at $t = \infty$, as
we did in Eqs.~(\ref{eq:E6}), (\ref{eq:F6}) and (\ref{eq:G6}).  On
the other hand, according to Eq.~(\ref{eq:M8}), we have
\begin{eqnarray}
&&\int^{\,t'}_{\!-\infty}G_0(x, x')\exp\big[i\omega(t -
t')\big]\,dt\nonumber\\
&&= \frac{\theta\big(k_F - |{\bf k}|\big)}{\omega -
\epsilon_0({\bf k})/\hbar - i\delta}\exp\big[i{\bf k\cdot(r -
r')}\big] \label{eq:M10}
\end{eqnarray}
Adding Eqs.~(\ref{eq:M9}) and (\ref{eq:M10}) together, we find
through Eq.~(\ref{eq:M7}) that
\begin{eqnarray}
&&\int^{\infty}_{\!-\infty}G_0(x, x')\exp\big[i\omega(t -
t')\big]\,dt\nonumber\\
&&= \frac{\exp\big[i{\bf k\cdot(r - r')}\big]}{\omega -
\epsilon_0\big({\bf k}\big)/\hbar + i\delta\,\mbox{sign}\big(|{\bf
k}| - k_F\big)}\label{eq:M11}
\end{eqnarray}
where sign($\xi$) = 1 if $\xi > 0$, otherwise sign($\xi$) = -1. It
is easy to recognize that Eq.~(\ref{eq:M11}) represents the
Fourier transform of $G_0(x, x')$ with respect to $t$. An inverse
Fourier transform of the expression on the right hand side of
Eq.~(\ref{eq:M11}) gives
\begin{eqnarray}
G_0(x, x') = \frac{1}{(2\pi)^4}\int d\omega\int d{\bf k}\;G_0({\bf
k}, \omega)\nonumber\\
\times\exp\big[i{\bf k\cdot(r - r')}\big]\exp\big[-i\omega(t -
t')\big] \label{eq:M12}
\end{eqnarray}
with
\begin{eqnarray}
G_0({\bf k}, \omega) = \frac{1}{\omega - \epsilon_0\big({\bf
k}\big)/\hbar + i\delta\,\mbox{sign}\big(|{\bf k}| - k_F\big)}
\label{eq:M13}
\end{eqnarray}
which is known as the  free electron Green function in frequency
and momentum space,


\begin{thebibliography}{9}

    \bibitem{ZW1}
    X. H. Zheng and D. G. Walmsley, J. Phys.: Condens. Matt. {\bf 16},
    8297 (2004).

    \bibitem{ZW2}
    X. H. Zheng and D. G. Walmsley, Phys. Rev. B {\bf 71}, 134512 (2005).

    \bibitem{BCS}
    J. Bardeen, L. N. Cooper and J. R. Schrieffer, Physs. Rev. {\bf
    108}, 1175 (1957).

    \bibitem{Abrikosov}
    A. A. Abrikosov, L. P. Gorkov and I. E. Dzyaloshinski, Methods
    of Quantum Field Theory in Statistical Physics (Dover, New YOrk,
    1975).

    \bibitem{Schrieffer}
    J. R. Schrieffer, Theory of Superconductivity (Persus, Reading,
    1983).

    \bibitem{Hubbard}
    J. Hubbard, Proc. Roy. Soc. (London) {\bf A240}, 539 (1957).


    \bibitem{Gell-Mann}
    M.\ Gell-Mann \& F.\ Low, Phys.\ Rev.\ {\bf 84}, 350 (1951).

    \bibitem{Wick}
    G. C. Wick, Phys. Rev. {\bf 80}, 268 (1950).

    \bibitem{Ziman}
    J. M.\ Ziman, Elements of advanced quantum theory (Cambridge
    University Press, 1969) 53.

    \bibitem{Frohlich}
    H. Fr\"ohlich, Proc.\ Roy.\ Soc.\ A {\bf 215}, 291 (1952).



\end{thebibliography}
\end{document}